\documentclass{article}
\begin{document}
\centerline{\Large{The Universe With Bulk Viscosity }}
\begin{center}
 Arbab I. Arbab\footnote{arbab@ictp.trieste.it}
 \\
{\small Department of Physics, Teacher's College,
Riyadh 11491, P.O.Box 4341, Saudi Arabia \\
Comboni College for Computer Science, P.O. Box 114, Khartoum,
Sudan }
\end{center}
\abstract{ Exact solutions for a model with variable $G$,
$\Lambda$ and bulk viscosity are obtained. Inflationary solutions
with constant (de Sitter-type) and variable energy density are
found. An expanding anisotropic universe is found to isotropize
during its course of expansion but a static universe is not. The
gravitational constant is found to increase with time and the
cosmological constant decreases
with time as $\Lambda \propto t^{-2}$.\\
    Variable $G$, $\Lambda$; Bianchi models;
Bulk viscosity; Inflation \\

%

\section{Introduction}           
\label{sect:intro} In a recent paper, Kalligas, Wesson and Everitt
(Kalligas {\it et al} 1995) have investigated a flat model with
variable gravitational ($G$) and cosmological $(\Lambda$)
``constants".
 In the same line, Singh {\it et al.} (Singh {\it et al} 1998) have considered a
 viscous cosmological model  with variable $G$ and $\Lambda$. They considered a
 different energy conservation  law from that of Arbab (Arbab 1997) but,
 they have found similar solutions as in  (Arbab 1997).
 Very recently, we have studied Bianchi type I model and we have shown that
 the universe isotropize during its course of expansion (Arbab 1998).
 With a similar approach, we wish to study the effect of anisotropy in the universe
 with the energy conservation advocated by Singh {\it et al.} where $G$ and $\Lambda$
 vary with time. We have shown that the introduction of bulk viscosity enriches this
 cosmology.\\
 Kalligas {\it et al.} have found solutions for a static universe with zero
 total energy density while $G$ and $\Lambda$ are allowed to vary with time.
 However, their solution, does not seem to be physically sensible, since one does
 expect  $G$ to vary with time in an empty universe !\\
 In the present case, we show that the static universe must be empty and has a vanishing
 cosmological constant ($\Lambda$) and bulk viscosity ($\eta$). We have also shown that the
 presence of viscosity helps an anisotropic universe to isotropize
 with expansion.  We have also obtained solutions with constant and variable
 energy density that correspond to either static or inflationary universe.
We remark that these solutions do not hold for our earlier work
(Arbab 1998). The gravitational constant is found to increase with
time. For a universe in balance (flat) expansion must accelerate
in order to overcome future collapse. The presently observed
acceleration of the universe is justified in the present models.
\section{SOLUTIONS FOR THE ISOTROPIC UNIVERSE}
\label{sect:sol} In a flat Robertson Walker metric
\begin{equation}
ds^2=dt^2-R^2(t)(dr^2+r^2d\theta^2+r^2\sin\theta^2d\phi^2)
\end{equation}
Einstein's field equations with time-dependent $G$ and $\Lambda$
read (Weinberg 1971)
\begin{equation}
{\cal{R}}_{\mu\nu}-\frac{1}{2}g_{\mu\nu}{\cal{R}}=8\pi
G(t)T_{\mu\nu}+\Lambda(t)g_{\mu\nu}
\end{equation}
The variation of the gravitational constant was first suggested by
Dirac (Dirac 1937) in an attempt to understand the appearance of
very large numbers, when atomic and cosmic worlds are compared. He
postulated that the gravitational constant ($G$) decreases
inversely with cosmic time. On the other hand, Einstein introduced
the cosmological constant ($\Lambda$) to account  for a stable
static universe, as appeared to him at that time. When he later
knew of the universe expansion he regretted its inclusion in his
field equations. Now cosmologists believe that $\Lambda$ is not
identically but very close to zero. They relate this constant to
the vacuum energy that first inflated our universe making it
expanding. A vacuum energy could correspond to, from a point view
of particle physics, a residue of some form of quantum field that
is diluted to its presently small value. However, other
cosmologists dictate a time variation of this constant in order to
account for its present smallness. The variation of this constant
could resolve some of the standard model problems. Like $G$ the
constant $\Lambda$ is a gravity coupling and both should therefore
be treated on an equal footing. A proper way in which $G$ varies
is incorporated in the Brans-Dicke theory (Brans \& Dicke 1961).
In this theory $G$ is related to a scalar field that shares the
long range interaction with gravity. In the literature $\Lambda$
takes several ansatz, like $\Lambda\propto R^{-2},\ \
\Lambda\propto H^2,\ \ \Lambda\propto 8\pi G\rho$, etc, and with
different reasoning (Overdin \& Cooperstock 1998 and Sahni \&
Starobinsky 1999). In the present work, we allow the variation of
$G$ to be cancelled by the variation of $\Lambda$ (Beesham 1986,
Abdel Rahman 1990, Pande 2000 Bonanno \& Reuter 2002). \\
Considering the imperfect-fluid energy momentum tensor
\begin{equation} T_{\mu\nu}=(\rho+p^*)u_\mu u_\nu-p^*g_{\mu\nu}
\end{equation}
eq.(2) yields the two independent equations
\begin{equation}
3\left(\frac{\ddot R}{R}\right)=-4\pi G(3p^*+\rho)+\Lambda\ \ ,
\end{equation}
and \begin{equation} 3\left(\frac{\dot R}{R}\right)^2=8\pi
G\rho+\Lambda\ .
\end{equation} Elimination of $\ddot R$ and the differentiated form of eq.(5) gives
\begin{equation} 3(p^*+\rho)\dot
R=-\left(\frac{\dot G}{G}\rho+\dot \rho+\frac{\dot\Lambda}{8\pi
G}\right)R\,
\end{equation}
where a dot denotes differentiation with respect to time $t$ and
$p^*=p-3\eta H$, $\eta$ being the coefficient of bulk viscosity,
$H$ the Hubble constant. The equation of state relates the
pressure ($p$) and the energy density ($\rho$) of the cosmic fluid
by the equation
\begin{equation}
p=(\gamma-1)\rho
\end{equation}
where $\gamma=\rm constant$. Vanishing of the covariant divergence
of the Einstein tensor in eq.(2) and the usual energy-momentum
conservation relation ($T^{\mu\nu}_{;\nu}= 0 $) lead to
\begin{equation}
 8\pi\dot G\rho+\dot\Lambda=0\ ,
\end{equation}
and
\begin{equation} \dot\rho+3(p^*+\rho)H=0,
\end{equation}
or
\begin{equation}
 \dot\rho+3(p+\rho)H=9\eta H^2\ .
\end{equation}
 We see that the bulk viscosity appears as a source term in the
energy conservation equation. Hence the RHS of eq.(10) would
correspond to the rate of thermal energy generated due to
viscosity. This may help solve the generation of entropy in the
universe associated with the standard model of cosmology.

In this paper we will consider the very special ansatzs (Arbab
1997)
\begin{equation}
\Lambda=3\beta H^2, \ \ \ \ \beta= \rm const.\ ,
\end{equation}
and
\begin{equation}
\eta=\eta_0\rho^n,\qquad \eta_0\ge 0,\ n=\rm const.
\end{equation}
We have shown very recently  that eq.(11) is equivalent to writing
$\Lambda$ as $\Lambda=\left(\frac{\beta}{\beta-3}\right)4\pi
G\rho$  for  a non-viscous model (Arbab 2002). This form is
interesting since it relates the vacuum energy directly to matter
content in the universe. Hence, any change in $\rho$  will
immediately imply a change in $\Lambda$, i.e. if $\rho$ varies
with cosmic time then $\Lambda$ also varies with cosmic time.

In what follows we will discuss the solution of the model
equations for an isotropic and an anisotropic universe with the
above prescription for $\Lambda$ and $\eta$.
\subsection{Solution with constant energy  density}
One can satisfy eq.(10) with a constant energy density ($\rho=\rm const)$ with:\\
(i) $H=0$, which implies a static universe.\\
(ii) $\eta=\eta_0\rho$ (i.e. $n=1$), $H=\frac{\gamma}{3\eta_0}=\rm
const.$. The solution of this equation is of the form $ R=\rm
const.\exp(Ht).$ We remark here that the classical inflation with
an equation of state $p=-\rho$ is not permitted in this model.
\subsection{Solution with variable energy density} \label{sect:constant}
{\it Inflationary solution with variable energy density}\\
Consider the bulk viscosity to have the form $\eta=\eta_0\rho$
(i.e. $n=1$). With  $H=H_0<\frac{\gamma}{3\eta_0}$\ we have $R=\rm
const.\exp(H_0t)$ so that eq.(10) yields a decaying mode of the
energy density given by
\begin{equation}
\rho=F\exp-3H_0(\gamma -3\eta_0 H_0)t\ ,\qquad F=\rm const..
\end{equation}
Now  consider the case $\beta=1$.\\ Equations (5), (8) and (11)
yield
\begin{equation} \Lambda=\rm const.\ ,
\qquad G=0
\end{equation}
We remark that this solution is not possible within the framework
of the conventional inflationary models. It is however remarked by
Abdel Rahman (Abdel Rahman 1990) that one possible interpretation
of eq.(14) is that  the universe came to being just prior to the
onset of gravity at $t=0$ as a result of a vacuum fluctuation
propelled by the repulsive effect of the positive cosmological
constant.\\

\section{SOLUTIONS FOR AN ANISOTROPIC UNIVERSE}
For the Bianchi type I metric
\begin{equation}
ds^2=dt^2-R_1^2dx^2-R_2^2dy^2-R_3^2dz^2
\end{equation}
with an imperfect-fluid energy momentum tensor, Einstein's field
equations yield (Arbab 1998)
\begin{equation}
\frac{\dot R_1\dot R_2}{R_1R_2}+\frac{\dot R_1\dot
R_3}{R_1R_3}+\frac{\dot R_2\dot R_3}{R_2R_3}=-8\pi G\rho-\Lambda\
,
\end{equation}
\begin{equation}
\frac{\dot R_1\dot R_2}{R_1R_2}+\frac{\ddot R_1 }{R_1}+\frac{\ddot
R_2}{R_2}=8\pi Gp^*-\Lambda\ \ ,
\end{equation}
\begin{equation}
\frac{\dot R_1\dot R_3}{R_1R_3}+\frac{\ddot R_1 }{R_1}+\frac{\ddot
R_3}{R_3}=8\pi Gp^*-\Lambda     \ \ ,
\end{equation}
\begin{equation}
\frac{\dot R_2\dot R_3}{R_2R_3}+\frac{\ddot R_2 }{R_2}+\frac{\ddot
R_3}{R_3}=8\pi Gp^*-\Lambda\ \ ,
\end{equation}
and
\begin{equation}
8\pi \dot G \rho+\dot \Lambda+8\pi
G\left[\dot\rho+(\rho+p^*)\left(\frac{\dot R_1}{R_1}+\frac{\dot
R_2}{R_2}+\frac{\dot R_3}{R_3}\right)\right]=0\ .
\end{equation}
where a dot denotes differentiation with respect to time $t$. From
eqs.(16)-(20) one obtains
\begin{equation}
\frac{\ddot R_1}{R_1}+\frac{\ddot R_2}{R_2}+\frac{\ddot
R_3}{R_3}=4\pi G(\rho+3p^*)-\Lambda \ .
\end{equation}
where $p^*=p-3\eta H$. The energy conservation
$(T^{\mu\nu}_{;\nu}=0$) implies that \begin{equation}
\dot\rho+3H(\rho+p^*)=0\ ,
\end{equation}
or
\begin{equation} \dot\rho+3H(\rho+p)-9\eta H^2=0\ .
\end{equation}
Here we define the average scale factor $R$ by $R\equiv
(R_1R_2R_3)^{1/3} $ so that
\begin{equation}
H=\frac{\dot R}{R}=\frac{1}{3}\left(\frac{\dot
R_1}{R_1}+\frac{\dot R_2}{R_2}+\frac{\dot R_3}{R_3}\right).
\end{equation} Using eqs.(22) and (24), eq.(20) yields
\begin{equation}
\dot\Lambda+8\pi\dot G\rho=0\ \ .
\end{equation}
Let us now assume that the energy density is given by the power
law
\begin{equation}
\rho=At^m  , \ \ \ A=\rm const.\ ,\ \ m =\rm\ const.\ ,
\end{equation}
and the average scale factor
\begin{equation}
R=Bt^\alpha  ,\ \  \ \alpha=\rm const.\ ,\ \ B= \rm const..
\end{equation}
Substituting  eqs.(26) and (27) in (23), (11) and (25) one gets
\begin{equation}
\rho=At^{-1/(1-n)} ,
\end{equation}
\begin{equation}
\Lambda=3\alpha\beta t^{-2}\ \ ,
\end{equation}
\begin{equation}
G =\frac{3\beta\alpha^2(1-n)}{4\pi A(2n-1)}t^{(2n-1)/(1-n)} \ ,\ \
n\ne \frac{1}{2}\ \ , \ 1\ .
\end{equation}
and the condition $m=\frac{-1}{1-n}$. For physical significance
$G>0$ so that $n > \frac{1}{2}$. This implies that the
gravitational constant is an ever-increasing function of time.
Consequently, for a flat (balanced) universe  the expansion must
increase (accelerate) so that the universe can remain in balance.
 Thus, the presently observed acceleration of the universe may be attributed
 to this ever growing gravity instead of invoking any exotic matter.\\
We now consider the anisotropy energy $(\sigma)$ defined by
\begin{equation}
8\pi G\sigma=\left(\frac{\dot R_1}{R_1}-\frac{\dot
R_2}{R_2}\right)^2+\left(\frac{\dot R_1}{R_1}-\frac{\dot
R_3}{R_3}\right)^2 +\left(\frac{\dot R_2}{R_2}-\frac{\dot
R_3}{R_3}\right)^2\ .
\end{equation}
Using eqs.(5), (16) and (24), the above equation becomes
\begin{equation}
8\pi G\sigma=18\left(\frac{\dot R}{R}\right)^2+48\pi
G\rho+6\Lambda\ .
\end{equation}
We see that the anisotropy energy ($\sigma$) becomes
\begin{equation}
8\pi G\sigma=Dt^{-2}, \ \ \ D=\rm \ const.\ ,
\end{equation}
a result that has been obtained by Arbab (Arbab 1998).
\subsection{Constant energy solution} \label{sect:constant energy}
Using eq.(12), eq.(23) can be written in the form
\begin{equation}
\dot\rho+3\gamma H\rho=9\eta_0\rho^n H^2\ .
\end{equation}
Now consider the following two cases:\\
(i) {\it Static universe} ( $H=0\ ,\ \ R=\rm \ const.$).\\
The above equation yields $\rho=\rm const$, and  eqs. (11) and
(25) yield  $\Lambda=0$ and $G=\rm const.$. It follows from eqs.
(5) and (32) that the  anisotropy energy $\sigma=0 $. It is shown
by Kalligas {\it et al.} ( Kalligas {\it et al.} 1995) that a
static universe can only be isotropic, i.e., $\sigma=0$ . However,
a static universe with constant energy density can not exist
unless $G=0$, as is evident from eq.(5). Hence, only an empty
static universe can exist. \\ We remark here that the claim made
by Kalligas {\it et al.} that a universe with a vanishing total
energy density has $G$ and $\Lambda$ vary with time is physically nonsensical.\\
(ii) {\it Inflationary universe with constant energy density}
($\rho=\rm const.\ ,\ $$H=\rm const.,\ n=1 $)\\
$H=H_0=\rm const.$ implies  $R=\rm const. exp(H_0t)$. Equation
(34) gives $H_0=\frac{\gamma}{3\eta_0}$\ .\\ Hence, eqs.(11), (25)
and (32) give $\Lambda=\rm const.$, $G=\rm const.$ and
$\sigma=\left(\frac{9H_0^2}{2\pi
G}\right)=\left(\frac{\gamma^2}{2\pi \eta_0^2G}\right)$, whether
$\Lambda=0$ or not.
\section{Conclusions}
\label{sect:conclusion} In this paper we have studied both
isotropic and anisotropic models with variable $G$, $\Lambda$ and
bulk viscosity. We have found that energy conservation is
guaranteed provided the three scalars, $G$, $\Lambda$ and $\eta$
conspire to satisfy it. We have found that a constant energy
density would lead to either a static or inflationary universe. We
have found that the classical inflation of an equation of state
$p=-\rho$ is not permitted. Inflationary solutions with constant
and variable energy density are found. These solutions are
influenced by the presence of the bulk viscosity.  Such a solution
is triggered by the presence of the bulk viscosity alone. An
initially expanding anisotropic universe is found to isotropize as
it evolves. However, a static empty universe must be isotropic.
This may explain why the present universe is very isotropic.
Finally, we have found that the gravitational constant to increase
with time in radiation-dominated $(\gamma=\frac{4}{3}$) and
matter-dominated ($\gamma=1$) epochs. The presently observed
acceleration of the universe may be interpreted as due to a
balance between expansion and gravity. And since gravity increases
expansion must accelerate!
\newpage
\subsection*{References}

Abdel Rahman, A.-M.M., (1990). Gen. Rel. Gravit. 22, 655.\\
Arbab, A.I., (1997).  Gen. Rel. Gravit. 29, 61.\\
Arbab, A.I., (1998). Gen. Rel. Gravit. 30, 1401. \\
Arbab, A.I., (2003). Class. Quantum Grav. 20, 93. \\
Beesham, A. (1986). IJTP, 25, 1295, (1986). Nouvo Cimento B96, 17.\\
Bonanno, A., and M. Reuter., (2002). xxx.arXiv.org/abs/astro-ph/0210472.\\
Brans, C., and Dicke, R.H., (1961). Phys. Rev. 124, 925.\\
Dirac, P.A.M., (1937). Nature 139, 323.\\
Kalligas, D., Wesson, D.P.S., and  Everitt, C.W.F. (1995). Gen. Rel. Gravit. 27, 645.\\
Overdin, J. M., and Cooperstock, F.I., 1998. Phys. Rev. D58, 043506.\\
Pande, H.D., {\it et al}., (2000). Indian J. pure appl. Math. 32,
161.\\
Sahni, V., and Starobinsky, A., 1999. Los Alamos preprint
astro-ph/9904398 \\
Singh, T., Beesham, A., and Mbokazi, W.S., (1998). Gen. Rel. Gravit. 30, 573.\\
Weinberg, S. (1971). {\it Gravitation and Cosmology} (Wiley, New York).\\
\label{lastpage}
\end{document}